\documentclass{acm_proc_article-sp}

\usepackage{amsmath}
\usepackage{amssymb}
\usepackage{graphicx}
\usepackage{float}
\usepackage{array}
\usepackage{booktabs}
\usepackage[ruled, vlined, linesnumbered]{algorithm2e}
\usepackage{algpseudocode}
\usepackage{multirow}

\makeatletter
\renewcommand\paragraph{%
   \@startsection{paragraph}{4}{0mm}%
      {-\baselineskip}%
      {.5\baselineskip}%
      {\normalfont\normalsize\bfseries}}
\makeatother

\DeclareMathOperator*{\argmax}{arg\,max}

\begin{document}

\title{R-Rec: A rule-based system for contextual suggestion using tag-description similarity}
%\subtitle{[Extended Abstract]
%\titlenote{A full version of this paper is available as
%\textit{Author's Guide to Preparing ACM SIG Proceedings Using
%\LaTeX$2_\epsilon$\ and BibTeX} at

\numberofauthors{3} %  in this sample file, there are a *total*
\author{
% 1st. author
\alignauthor
Kshitij Singh\\
       \affaddr{Department of Computer Science and Engineering}\\
       \affaddr{Indian Institute of Technology (BHU)}\\
       \affaddr{Varanasi, India}\\
       \email{\small kshitij.singh.cse12@iitbhu.ac.in}
\alignauthor
Manajit Chakraborty\titlenote{Corresponding Author}\\
			 \affaddr{Department of Computer Science and Engineering}\\
       \affaddr{Indian Institute of Technology (BHU)}\\
       \affaddr{Varanasi, India}\\
       \email{\small cmanajit.rs.cse14@iitbhu.ac.in}
% 2nd. author
% 3rd. author
\alignauthor C Ravindranath Chowdary\\
       \affaddr{Department of Computer Science and Engineering}\\
       \affaddr{Indian Institute of Technology (BHU)}\\
       \affaddr{Varanasi, India}\\
       \email{\small rchowdary.cse@iitbhu.ac.in}
%\and  % use '\and' if you need 'another row' of author names
%% 4th. author
%\alignauthor Lawrence P. Leipuner\\
       %\affaddr{Brookhaven Laboratories}\\
       %\affaddr{Brookhaven National Lab}\\
       %\affaddr{P.O. Box 5000}\\
       %\email{lleipuner@researchlabs.org}
%% 5th. author
%\alignauthor Sean Fogarty\\
       %\affaddr{NASA Ames Research Center}\\
       %\affaddr{Moffett Field}\\
       %\affaddr{California 94035}\\
       %\email{fogartys@amesres.org}
%% 6th. author
%\alignauthor Charles Palmer\\
       %\affaddr{Palmer Research Laboratories}\\
       %\affaddr{8600 Datapoint Drive}\\
       %\affaddr{San Antonio, Texas 78229}\\
}
% There's nothing stopping you putting the seventh, eighth, etc.
% author on the opening page (as the 'third row') but we ask,
% for aesthetic reasons that you place these 'additional authors'
% in the \additional authors block, viz.
%\additionalauthors{Additional authors: John Smith (The Th{\o}rv{\"a}ld Group,
%email: {\texttt{jsmith@affiliation.org}}) and Julius P.~Kumquat
%(The Kumquat Consortium, email: {\texttt{jpkumquat@consortium.net}}).}
% Just remember to make sure that the TOTAL number of authors
% is the number that will appear on the first page PLUS the
% number that will appear in the \additionalauthors section.

\maketitle
\begin{abstract}
Contextual Suggestion deals with search techniques for complex information needs that are highly focused on context and user needs. In this paper, we propose \emph{R-Rec}, a novel rule-based technique to identify and recommend appropriate points-of-interest to a user given her past preferences. We try to embody the information that the user shares in the form of rating and tags of any previous point(s)-of-interest and use it to rank the unrated candidate suggestions. The ranking function is computed based on the similarity between a suggestion and the places that the user like and the dissimilarity between the suggestion and the places disliked by the user. Experiments carried out on TREC-Contextual Suggestion 2015 dataset reveal the efficacy of our method.
\end{abstract}

%\begin{CCSXML}
%<ccs2012>
 %<concept>
  %<concept_id>10010520.10010553.10010562</concept_id>
  %<concept_desc>Computer systems organization~Embedded systems</concept_desc>
  %<concept_significance>500</concept_significance>
 %</concept>
 %<concept>
  %<concept_id>10010520.10010575.10010755</concept_id>
  %<concept_desc>Computer systems organization~Redundancy</concept_desc>
  %<concept_significance>300</concept_significance>
 %</concept>
 %<concept>
  %<concept_id>10010520.10010553.10010554</concept_id>
  %<concept_desc>Computer systems organization~Robotics</concept_desc>
  %<concept_significance>100</concept_significance>
 %</concept>
 %<concept>
  %<concept_id>10003033.10003083.10003095</concept_id>
  %<concept_desc>Networks~Network reliability</concept_desc>
  %<concept_significance>100</concept_significance>
 %</concept>
%</ccs2012>  
%\end{CCSXML}
%
%\ccsdesc[500]{Computer systems organization~Embedded systems}
%% End generated code
%%
%
%%
%%  Use this command to print the description
%%
%\printccsdesc

% We no longer use \terms command


\section{Introduction}
Recommending items to users based on her personal preferences and choices has been a long standing problem. With the emergence of social networks and aggressive expansion of e-commerce, this field has attracted serious attention. One of the many variants of recommender systems include contextual suggestion. Contextual suggestion aims to make appropriate points-of-interest recommendations to a traveler traveling to an unfamiliar city, given her past preferences \cite{Dean15}. The premise of the problem is challenging because users (in this instance travelers) come from various backgrounds and their needs and taste vary widely. This problem is sometimes compounded when the user is reluctant in sharing her opinion of some particular point(s)-of-interest or attraction that she has visited earlier. Moreover, it might sometimes be difficult to judge an attraction based on the limited amount of information that is available on the web. \\For ease of understanding let us consider an example. Say, Radha is a person living in India who loves to travel. Her past travel experiences include Goa, Darjeeling, Kochi etc. Let us assume that we are given a snapshot of the attractions or points-of-interests that she has visited over past trips. Given that we are provided with some background knowledge about her tastes and choices, would it be possible for us to suggest her new point-of-interests at a new place? Let us first take a look at her choice of attractions from past trips (Table \ref{tab:eg}).\\
\begin{table}[h!]
\centering
\scriptsize
\begin{tabular}{|p{1.5cm}|p{2.5cm}|c|p{3cm}|} \hline
Place& Attraction & Rating & Tags\\ \hline\hline
\multirow{4}{4pc}{Goa}& Sahakari Spice Farm&6/10&Foodie, Peace, Nature Lover, Beach \\\cline{2-4} 
										&Tomb of St. Francis Xavier&7/10&Peace, History Buff, Art \& Architecture\\\cline{2-4}
										&Panjim&8/10&Foodie, Beach Goer, Nature Lover, Peace\\\cline{2-4}
										&Club Cabana& 5/10 & Beach Goer, Like a Local, Thrill Seeker, Nightlife Seeker\\\hline
\multirow{4}{4pc}{Kochi} & Wonderela Amusement Park & 7/10 & Theme Parks, Water, Amusement\\ \cline{2-4}
															 & Folklore Musuem & 9/10 & History Museums, Art \& Architecture, History Buff\\ \cline{2-4}
															 & Santa Cruz Basilica & 8/10 & Sights \& Landmarks, Architectural Buildings\\ \cline{2-4}
															 & LuLu Mall & 6/10 & --\\ \hline
\multirow{4}{4pc}{Darjeeling} & Kanchenjunga Mountain & 10/10 & Mountains, Nature \& Parks\\ \cline{2-4}
														& Padmaja Naidu Zoological Park & 8/10 & Outdoor Activities, Zoos \& Aquariums, Nature \& Parks\\ \cline{2-4}
														& Peace Pagoda &--&--\\ \cline{2-4}
														& Passenger Ropeway & 8/10& Tramways, Thrill Seeker, Transportation\\ \hline

\end{tabular}
\caption{Snapshot of places visited by Radha}
\label{tab:eg}
\end{table}
Now suppose she is planning to take a trip to Shimla. Assuming that the above Table \ref{tab:eg} information is available with the system can it make proper recommendations to suit her needs? While for a human being it is easier to decipher what Radha likes and dislikes, for a computer to do so is certainly cumbersome. In such cases, rule-based systems could come in handy. For example, from the above table we see that Radha is inclined towards peaceful environments followed by architectural landmarks, historical sites etc. If we assume that 7/10 is her par rating and anything below that is something she didn't like much we can set up some rules for attractions so that points-of-interest with her likings and ratings are suggested to her on the top of the list. Considering such constraints, a recommender system can then suggest her the following attractions at Shimla (Table \ref{tab:sugg}).\\
\begin{table}[!h]
\centering
\begin{tabular}{|c|p{3cm}|}\hline
Attraction& Tags\\ \hline \hline
Viceregal Lodge& Architectural Buildings, Sights \& Landmarks\\ \hline
Shimla Christ Church & Sights \& Landmarks, Architectural building\\ \hline
Gaiety Heritage Cultural Complex & Historical landmark, Art \& Architecture\\ \hline
Kalka-Shimla Railway &  Scenic Railroads, Tours, Peace \\ \hline
\end{tabular}
\caption{Sample of recommendation list}
\label{tab:sugg}
\end{table}

The TREC Contextual Suggestion track was started in 2012 with the aim of investigating complex information needs that are highly dependent on context and user interests. The problem was modeled as a travel-recommendation problem where given some user profiles and contexts (containing information about points-of-interest), the participants had to suggest attractions to the users visiting a new city. Over the years, the track has updated and modified their guidelines as to how those contexts could be utilized and the platform for participation (live or batch experiment). In 2015, the contexts consisted of a mandatory city name which represents which city the trip will occur in and several pieces of optional data about the trip (such as trip type, trip duration, season \emph{etc.}). Instead of providing the detailed information about each attraction, they were accompanied with a URL. Consequently, the data had to be extracted by participants from open-web for processing. We have used the dataset provided in this track and evaluated our procedure using the same metrics (P@5 and MRR) that were used by TREC to evaluate its participants. Our experimental results show that our rule-based system outperforms all other systems in both the metrics.
\section{Related Work}
The advent of e-commerce has revolutionized many industries including tourism \cite{Werthner04}. With better reach in internet connectivity and easy access to portable devices, online travel bookings have seen an upward surge along with online itinerary planning \cite{Haubl04}. Since recommendation systems are perceived as one of the fastest growing domains of internet applications, its footprint in tourism industry is also evident \cite{Fesenmaier06}. Designing recommender systems for the tourism industry is challenging as a travel-related recommendation must refer to a variety of aspects such as locations, attractions, activities \emph{etc.} to provide a meaningful suggestion. Also, the recommender system needs to keep a record of the past history of users (which can be done either implicitly or explicitly) and analyze the same \cite{Bobadilla13}. \\Context-aware recommendation has been successfully applied in various forms to suggest items to users across domains. Travel recommendation is no exception to that. But, inherently there is a catch to it since usage of too many contextual variables may lead to a drastic increase in dimensionality and loss of accuracy in recommendation. Zheng \emph{et al.} \cite{Zheng12} tackles this problem by decomposing the traditional collaborative filtering into three context-sensitive components and then unifying them using a hybrid approach. This unification involves relaxation of contextual constraints for each component. The problems and challenges of tourism recommendation systems have discussed in \cite{Berka04} along with their potential applications. A detailed survey of travel recommendation systems covering aspects like techniques used, functionalities offered by each systems, diversity of algorithms \emph{etc.} can be found in \cite{Borras14}.\\
Tourists today have complex, multi-faceted needs and are often flexible, experienced and demand both perfection and diversity in recommendation \cite{Ricci02}. To deal with such vagaries, travel recommendation systems today rely on latest technological offerings such as GPS \cite{Zheng09}\cite{Zheng10}, geo-tagged images \cite{Kurashima10}\cite{Majid13} and community contributed photos \cite{Cheng11}. Choi \emph{et al.} \cite{Choi09} has suggested the use of travel ontology for recommending places to users. Sun \emph{et al.} \cite{Sun15} build a recommendation system that provides users with the most popular landmarks as well as the best travel routing between the landmarks based on geo-tagged images.\\
Portable devices such as mobile phones and tablets provide information which can be used to recommend travel-related multimedia content, context-aware services, ratings of peers \emph{etc.} \cite{Park07} \cite{Gavalas11}. A systematic study on mobile recommender systems can be found in \cite{Gavalas14}.  

\section{Proposed Approaches}
In this section we describe the various approaches adopted for recommending the best set of results to the user based on his/her context. Although we will show in a later section that \emph{R-Rec} turns out to be the best system for the current scenario, we give a brief description of other approaches that led to R-Rec. 
\subsection{Experimental Setup}
\subsubsection{Dataset}
 For experimental purposes, we used the dataset provided as part of the shared task in TREC 2015 Contextual Suggestion track.
\subsubsection{Preprocessing}
Both the candidate suggestions and user preferences (contexts) had URLs of the attractions. To get more information about the point-of-interests we initially crawled the webpages using a script but found that many of the links were either broken or dead. As an alternative measure, we sought the help of various publicly available APIs such as Alchemy API, Yelp API and Foursquare API to get more details about the attractions. Of all the three APIs, Foursquare proved to be the most useful one because the content it provided was rich as well as concise. Even after doing so, there were some residual websites for which the APIs failed to fetch any information. In such cases, we used whatever information we could crawl out of the webpages. Such contents were curated by removing stopwords and unintelligible characters. Also for convenience we retained only nouns, prepositions and adjectives from such content.
\subsubsection{Profile enrichment} \label{sec:tagging}
It is interesting to note that user assigned tags play an important role in determining how close our recommendation is to user's need. Taking a cue from this, we tagged the user preference attractions with tags from the predefined tag set that matched the description of URLs. For example, say user \textit{X} has 32 user preferences \emph{i.e.} attractions that she has visited previously. Now, it may so happen that she might feel reluctant to assign tags to all of them. Say, \textit{X} has rated all 32 attractions but tagged only 7 of them. So essentially, for those 25 attractions we have no idea what she liked or disliked about the place. The ratings might reflect how satisfied she was with the facilities at a particular attraction, but it does not reveal her qualitative judgement of that attraction.\\
To incorporate this fact, we extracted information of the user visited attractions from the URLs and matched them against the complete set of tags. Hence, an attraction which matched with say \textit{beach} or \textit{beach walk} was tagged with the tag ``beach''. Now while this was a naive approach this introduced a dilemma-- \textit{Which tags would be appropriate for a particular attraction?} It is evident from the dataset that almost all attractions have been tagged with multiple tags. Because it is customary that an attraction which is tagged with ``Restaurant'' should most certainly contain the tag ``Food''. While it is up to the users, whether they tag an attraction or not, this kind of incomplete information is certainly detrimental to our recommender system.\\
To tackle this scenario, we utilized WordNet \cite{Miller95}, from which we extracted synonyms for each tag and created a list (synset).  Each element in the synset of each of the tags was matched against the descriptions, and if any of the synonyms matched with high accuracy (similarity greater than $0.75$), we assigned that particular attraction with the original tag present in the provided tag set. For example, if any of the descriptions contained a word \emph{meal}, it was tagged with ``Food'' since `meal' is a synonym of `food'. It is to be noted that in the case of bi-gram or tri-gram tags each of the terms present in the tags were expanded using WordNet simultaneously.
This exercise was repeated for all the user visited attractions until there were no untagged attractions\footnote{\scriptsize It should be noted that we did not assign any new tags for attractions having user provided tags.}.
\subsubsection{Evaluation}
The evaluation of our approaches was performed using the relevance judgment and evaluation scripts provided for the batch task by TREC. The metrics used were Precision at k=5 (P@5) and Mean Reciprocal Rank. In this evaluation, a suggestion is relevant if it is rated 3 or 4 by user. Each of the approaches were compared against the TREC 2015 Contextual Suggestion (CS) batch experiment results median. \emph{T-Rec} was additionally compared against the best system available at TREC-CS 2015.
\subsection{D-Rec}
User profiles contain tags for various attractions along with ratings. These tags are essential in capturing the user's taste because it gives us a window view into why she liked or disliked that particular attraction. At the same time, it is also obvious that the description of user preference attractions will give us more information about the attraction. We hypothesized that appropriate selection of either measure would result in a better recommendation list. With this in mind, we computed two scores for a candidate suggestion-- one for user assigned tags to attractions $\mathcal{S}_c^t$ and the other for description of user visited attraction $\mathcal{S}_c^d$.
\subsubsection{Methodology}
\paragraph{Computing $\mathcal{S}_c^d$}
As stated earlier, we had extracted descriptions of both candidate suggestion attractions and user profile attractions using various APIs and scripts. For computing $\mathcal{S}_c^d$, the candidate suggestion attraction description is matched against each of the user profile attraction description and the matching score is calculated using the following formula:
\begin{center} \begin{equation} \label{eq:1}\mathcal{S}_{c_j}^d=\frac{similarity(d_{u_{ix}},d_{c_j})}{|d_{u_ix}||d_{c_j}|} \end{equation}\end{center}
where $d_{u_{ix}}$ and $d_{c_j}$ stands for profile attraction description and candidate suggestion description respectively. The denominator represents the product of the number of words present in each of the descriptions $d_{u_{ix}}$ and $d_{c_j}$ respectively. The similarity measure used here was WordNet similarity\footnote{\tiny http://search.cpan.org/tpederse/WordNet-Similarity-2.07/lib/WordNet/Similarity.pm}.
\paragraph{Computing $\mathcal{S}_c^t$}
Now, for calculating scores $\mathcal{S}_c^t$, we consider the profile attraction tags as a single phrase and use a similar formula as in Equation \ref{eq:1}.
\begin{center} \begin{equation} \label{eq:2}\mathcal{S}_{c_j}^t=\frac{similarity(<t_{u_i}>,d_{c_j})}{|<t_{u_i}>||d_{c_j}|} \end{equation}\end{center}
where $<t_{u_i}>$ denotes the tags of a profile attraction $u_i$ being considered as a phrase. $|<t_{u_i}>|$ denotes the number of words in the phrase $<t_{u_i}>$.
\paragraph{Ranking of candidate suggestions}
We assume that the tags provided by users in their profiles play a definitive role in describing user's choice and taste. Consequently, it must have higher priority than the description extracted from the attraction's website. Because in general, an attraction's website will contain information about the type, facilities and services that are provided by them. This may very well differ from how the user perceives that attraction and her experience of the same. Adhering to this fact, we use $\mathcal{S}_{c_j}^t$ to rank the candidate suggestions. In case, there is a tie it is resolved in favor of a candidate suggestion having higher $\mathcal{S}_{c_j}^d$. The complete procedure is summarized in Algorithm \ref{algo:att1}.\\
\begin{algorithm}[h]
\SetAlgoLined
 \KwData{Tagged user preferences $U$, Candidate suggestions $C$, Suggestion descriptions $d_C$}
 \KwResult{Recommendation list $R$}
 \caption{D-Rec}
\label{algo:att1}
\For{$1\leq i \leq |U|$}{
 \For{$1\leq j \leq |C|$}{
 
	\For{$1\leq x \leq |u_i|$}{
 $\mathcal{S}_{c_j}^d\leftarrow \frac{similarity(d_{c_j},d_{u_{ix}})}{|d_{u_{ix}}||d_{c_j}|}$\;
 
\For{$\forall t_k \in u_i$}{
$append$($<t_{u_i}>$,$t_k$)\;
}
 $\mathcal{S}_{c_j}^t\leftarrow \frac{similarity(d_{c_j},<t_{u_i}>)}{|<t_{u_i}>||d_{c_j}|}$\;
 }
}
}
$R\leftarrow Sort_{\mathcal{S}_{c_j}^t}(C)$\Comment{\textrm{\small resolve ties using }$\mathcal{S}_{c_j}^d$}\;
\end{algorithm}
The \emph{Sort()} function in Algorithms \ref{algo:att1} through \ref{algo:rrec} sorts the list of suggestions based on some score and returns the sorted list.
\subsubsection{Results and Analysis}
In Table \ref{tab:desc}, we present the results of this approach.
\begin{table}[!ht]
\centering
\caption{Evaluation results of D-Rec}
\label{tab:desc}
\begin{tabular}{|c|c|c|}\hline
System &P@5 & MRR\\ \hline \hline
Median of TREC-CS 2015 \cite{TREC15}& 0.5090 & 0.6716\\ \hline
D-Rec& 0.4701 & 0.6051\\ \hline
\end{tabular}
\end{table}
Clearly, this approach did not yield satisfactory results. The reason for this system's failure could be ascribed to the naive matching of tags against descriptions. While WordNet \emph{similarity} tends to capture semantic similarity to a certain extent, it is quite possible for it to miss on few terminologies and inter-word relationships. This could have resulted in a poor overall score. So, in the rest of the approaches we use direct string matching instead of WordNet similarity. Another drawback of this approach is that it is expensive in terms of time complexity because of matching $m\times n$ combinations where $m$ and $n$ are the lengths of any two descriptions respectively.
\subsection{C-Rec}
When the previous attempt of matching descriptions against tags did not yield decent results, we resorted to matching tags against tags. The possible reason for \emph{D-Rec}'s under-performance was that matching long descriptions was diminishing the overall score. But, instead if we had tags for the candidate suggestions like user profile attractions, intuitively the comparison and agreement between two sets of tags would be better. Thus, we first followed a similar procedure of tagging candidate suggestion descriptions as described in Section \ref{sec:tagging}. Thus, each candidate suggestion now consists of a set of tags from the available tag set. We propose two measures for capturing user's preference namely \emph{Coverage} and \emph{Completeness}.
\subsubsection{Coverage}
Our first step was to ensure that our recommendation should be able to satisfy user's diverse needs as stated in \cite{He14}. Tags were our only source of information about the facilities and services user is looking for in an attraction. Hence, for each user profile we compiled a set of non-redundant tags (denoted by $\tau_{u_i}$) from tags of all the visited attractions. Now, tags of each of the candidate suggestions $c_j$ are matched against all the tags in $\tau_{u_i}$. The number of matched tags are divided by the total number of tags in $\tau_{u_i}$, to generate a normalized coverage score $\theta_{c_j}$ for candidate suggestion $c_j$. Coverage ensures that the results in the recommended list embody most of the user specified needs. $\theta$ is computed as :
\begin{center}\begin{equation} \label{eq:3} \theta_{c_j}=\frac{|{\tau_{u_i}\cap \mathbf{t}_{c_j}|}}{|\tau_{u_i}|}\end{equation}\end{center}
where $\mathbf{t}_{c_j}$ stands for the set of tags pertaining to a candidate suggestion $c_j$.
\subsubsection{Completeness}
To ensure that along with coverage, individual type of needs are also catered to, matching of the candidate suggestion $c_j$'s tags with each of the taglist of user profile attractions is carried out. The sum of tags matching across all the profile attractions is divided with the total number of tags present in that user profile. This gives us a normalized score $\omega_{c_j}$ representing the \textit{completeness} of $c_j$. \emph{E.g.,} let us consider that there are five tags for some $c_j$, out of which $p$ tags match with profile attraction $i$, $q$ tags match with profile attraction $i+1$ and so on. So, if we sum the number of matched tags ($p+q+$...) and normalize it over total number of tags across profile attractions, we get an idea of how relevant the candidate suggestion is to the user profile. $\omega_{c_j}$ can be expressed in the form of Equation \ref{eq:4}.
\begin{center} \begin{equation}\label{eq:4} \omega_{c_j}=\frac{\sum_{i}{|\mathbf{t}_{u_i}\cap \mathbf{t}_{c_j}|}}{\sum_{i}|\mathbf{t}_{u_i}|}\end{equation}\end{center}
Here, $\mathbf{t}_{u_i}$ denotes the complete tag list for a user profile attraction $u_i$.
\subsubsection{Ranking of suggestions}
The scores calculated using Equations \ref{eq:3} and \ref{eq:4} are then used to rank candidate suggestions. We formulate two alternate approaches to generate the recommendation list. In first approach termed as \emph{Cov-Rec}, we give more preference to coverage over completeness. The other approach \emph{Cmp-Rec} is \textit{vice-versa} of \emph{Cov-Rec}. In either case, the non-preferred score is used to resolve ties among candidates. The reason behind incentivizing coverage in \emph{Cov-Rec} is that we want to cover as many user specified facilities as possible. The argument for favoring completeness over coverage in \emph{Cmp-Rec} is that we want to cater to important needs rather than serving all interests of a user. For either case, Algorithm \ref{algo:att2} states the generalized procedure \emph{C-Rec}.
\begin{algorithm}[!h]
\SetAlgoLined
 \KwData{Tagged user preferences $U$, Candidate suggestions $C$}
 \KwResult{Recommendation list $R$}
 \caption{C-Rec}
\label{algo:att2}
 \For{$1\leq i \leq |C|$}{
\For{$\forall u_k \in U$}{
$\tau_{u_k}\leftarrow \cup t_{k}$\;
$\theta_{c_i}^k$ \small is computed using Equation \ref{eq:3} \;
}
$\omega_{c_i}$ \small is computed using Equation \ref{eq:4}\; %=\frac{\sum_{j}\mathbf{t}_{u_j}\cap \mathbf{t}_{c_i}}{\sum_{j}|\mathbf{t}_{u_j}|}$
}
$R \leftarrow Sort(C)$ \Comment{\scriptsize based on either $\omega_{c_i}$ or $\theta_{c_i}$}\;
\end{algorithm}

\subsubsection{Result and Analysis}
Table \ref{tab:tag} presents the results obtained with \emph{C-Rec}.
\begin{table}[!ht]
\centering
\caption{Evaluation results of C-Rec }
\label{tab:tag}
\begin{tabular}{|c|c|c|}\hline
System &P@5 &MRR\\ \hline \hline
Median of TREC-CS 2015 \cite{TREC15}& 0.5090 & 0.6716\\ \hline
Cov-Rec & 0.5346 & 0.6735\\ \hline
Cmp-Rec & 0.5441 & 0.6839\\ \hline
\end{tabular}
\end{table}
In this case, although there is a noticeable improvement over \emph{D-Rec} there is still some scope for improvement. But an interesting observation is that preferring completeness over coverage gives better results than \textit{vice-versa}. Thus, it is imperative that major user needs be identified and catered to for suggesting the best recommendation to the user.\\
Unlike \emph{D-Rec}, this method establishes that tags are essential in capturing user's wishes and demands. It is to be noted that in both these attempts (\emph{D-Rec} and \emph{C-Rec}), we have not considered ratings for generation of recommendation list. This was based on the fact that user's rating may not inform us of her intention. She may very well be interested in the facilities offered by a kind of attraction. But, it may be the case that at some particular attraction, her satisfaction levels (pertaining to those facilities) weren't met and consequently she rated it low.
\subsection{R-Rec}
\subsubsection{Tag enrichment}
All the while we have been experimenting keeping in mind the requirements of the user under consideration. One interesting thing to note is the fact we have ignored that user assigned tags play an important role in determining how close our recommendation is to user's need. Taking cue from this, in this approach we tagged the user preference attractions with tags from the tagset that matched with the description of URLs. For example say user X has 32 user preferences \emph{i.e.} attractions that she has visited previously. Now, it may so happen that she might feel reluctant to assign tags to all of them. Say, she has rated all 32 attractions but tagged only 7 of them. So essentially, for those 25 attractions we have no idea what she liked or disliked about the place. The ratings might reflect how satisfied she was with the facilities at a particular attraction, but it doesn't reveal her qualitative judgement of that attraction.\\
To incorporate this fact, we extracted information of the user visited attractions from the URLs and matched them against the complete set of tags. Hence, an attraction which matched with say \textit{beach} or \textit{beach walk} was tagged with the tag ``beach''. Now while this was a naive approach this introduced a dilemma-- \textit{Which tags would be appropriate for a particular attraction?}. It is evident from the dataset that almost all attractions have been tagged with multiple tags. Because it is customary that an attraction which is tagged with ``Restaurant'' should most certainly contain the tag ``Food''. While it is upto users whether she has tagged an attraction or not, this kind of incomplete information is certainly detrimental to our recommender system.\\
To tackle this scenario, we reverted to WordNet \cite{Miller95}, from which we extracted synonyms for each tag and created a list for each tag. Another observation was that some of the tags were generic like \textit{Food} which could alternatively be expressed with the tag \textit{Cuisine}. On the other hand, some tags were very specific like ``Shopping for shoes''. Consequently, using a uniform tag matching approach would very likely result in wrong tagging of the attractions. To avoid such mistakes, we manually curated the tags into two categories-- 
\begin{itemize}
\item Generic Tags
\item Specific Tags
\end{itemize}
Matching of specific tags was simple. We matched the descriptions with the tag directly using WordNet's \textit{wup\_similarity}. If the similarity score was above 0.95 (to ensure a high degree of matching), we assigned that particular attraction with that specific tag. Again, each element in the synset of each of the generic tags were matched against the descriptions and if any of the synonyms matched with high accuracy we assigned that particular attraction with the generic tag present in the original tagset. For example, if any of the description contains a word \emph{meal}, it was tagged with ``Food'' since `meal' is a synonym of `food'. It is to be noted that in case of bi-gram or tri-gram tags each of the terms present in the tags were expanded using WordNet simultaneously.
This exercise was repeated for all the user visited attractions until there were no untagged attractions\footnote{\scriptsize It should be noted that in we left those attractions untouched for which user had already provided tags.}.
\subsubsection{Tag-matching and ranking of suggestions}
Once the above procedure of tagging user preferences is over, we move on to the most important part of the algorithm-- scoring each candidate suggestion and ranking them. 
\paragraph{Computing individual tag scores}
There is always a trade-off when it comes to giving importance to either ratings or tags for attractions visited by users. For our consideration we consider rating as the primary factor, because we hypothesize that a user will only rate an attraction X higher if he likes that particular kind of attraction and was satisfied with the service or facilities available at X. Keeping this in mind we assign scores to each tag pertaining to each rating for each user. The steps below summarize the procedure:
\begin{enumerate}
\item For each of the available ratings in the user preference list (context) we create an empty list $l$.
\item Initially for all tags occurring in a particular user context we assign a score of zero \emph{i.e.} \textit{Food} is assigned 0, \textit{Park} is assigned 0 and so on.
\item Now, it is quite common for users to rate two or more attractions with same ratings. In that case, it is important that we suggest the candidate items to users with equal ratings if they are on higher side. So, for each rating the total number of tags associated with it are computed and assigned to the previously empty list $l[r_i]$. Essentially it creates a list of lists like in Table \ref{tab:rate}.
\begin{table}[!h]
\centering
\begin{tabular}{|c|c|}\hline
Rating& Count of Tags\\ \hline \hline
4& 13\\ \hline
3&17\\ \hline
2&9\\ \hline
1&11\\ \hline
0&2\\ \hline
-1&0\\ \hline
\end{tabular}
\caption{List of tags for each rating}
\label{tab:rate}
\end{table}
\item To calculate the normalized score of each tag $\hat{s}(t_k^{r_i})$ pertaining to a rating $r_i$, we normalize the count of a particular tag $count_{r_i}(t_k)$ occurring in that rating by dividing it with the total count of tags for that particular rating $l[r_i]$ where $r_i\in r=\{4,3,2,1,0,-1\}$. \emph{E.g.} if the tag ``Food'' occurs four times among all the user preferences having rating 3, then the normalized score for ``Food'' at rating 3 is $\frac{4}{17}$.\\ This step ensures that the sum of scores of all tags for a particular rating is equal to one\footnote{\scriptsize $\sum_{n}s(t_n^{r_i})=t_1^{r_i}+t_2^{r_i}+...+t_n^{r_i}=1$}, so that no bias is induced for any rating. Lines 2-13 of Algorithm \ref{algo:rrec} summarizes the above steps.
\end{enumerate}
\paragraph{Scoring and ranking of candidate suggestions}
Once score for each tag for a user context has been computed we match each of the candidate suggestion tags with the rated attraction tags. Initially, all the candidate suggestions are assigned a score $\mathcal{S}_{c_j}=0$. For each suggestion, if any of the rated attraction's tag matches with candidate's, then its normalized score is added to the total score of the candidate suggestion. Mathematically,
\begin{center}\begin{equation} \mathcal{S}_{C_j}=\sum_{1\leq k \leq m}\hat{s}(t_k^{r_i}) \end{equation}\end{center}
where $m$ is the number of matched tags for a particular candidate suggestion $j$.\\
This process of matching a candidate suggestion is carried out for each of the user preferences. Let the rating user preference for which a candidate suggestion $j$'s score is maximum be $r$. Then, that particular candidate $j$ is assigned a rating of $r$ as well along with the candidate score of  $\argmax_S \mathcal{S}_{C_j}$. Based on this candidate score, the candidate suggestions are ranked in descending order. The candidates with highest ratings are kept on top. If there are multiple candidates with same rating then they are sorted on the basis of their individual score $\mathcal{S}_c$. If any candidate suggestion's tags didn't match with any of the user preference attraction tags, then it is assigned a score of zero. This process is reflected in lines 14-26 of Algorithm \ref{algo:rrec}.
\begin{algorithm}[h]
\SetAlgoLined
 \KwData{Tagged user preferences $U$, Rating $r$, Tagged candidate suggestions $C$, Empty list $l$}
 \KwResult{Recommendation list $R$}
 \caption{R-Rec}
\label{algo:rrec}
 %tf=
\For{$1\leq j \leq |U|$}{
 \For{$\forall r_i \in U_j$}{
 $l[r_i]\leftarrow 0$ \Comment{$r_i\in r=\{4,3,2,1,0,-1\}$}\;
 }
\For{$\forall r_i \in U_j$}{
\If{$t_k \in tagset[U_{j_x}] \land rating[U_{j_x}]=r_i$}{
$l[r_i]\leftarrow l[r_i]+1$ \Comment{\textrm{\scriptsize $1\leq x \leq |U_j|$,$1\leq k \leq |t_{U_j}|$}}\;
}}
\For{$\forall t_k \in r_i$}{
$s(t_k^{r_i})\leftarrow count_{r_i}(t_k)$\;
$\hat{s}(t_k^{r_i})\leftarrow \frac{s(t_k^{r_i})}{l[r_i]}$\;
}
\For{$\forall C_i \in C$}{
\For{$\forall U_{j_x} \in U_j$}{
\If{$\exists t_{C_i}\in match(t_{U_{j_x}},t_{C_i})=True$}{
$\mathcal{S}_{C_i}^x\leftarrow \sum_{m} \hat{s}(t_{U_{j_x}}^r)$ \Comment{\textrm{\scriptsize m = |matched tags|}}\;
}
\Else{
$\mathcal{S}_{C_i}^x\leftarrow 0$\;}
}
$\mathcal{S}_{C_i}\leftarrow \argmax_S \mathcal{S}_{C_i}^x$\;
$rating[C_i]\leftarrow rating[\argmax_S \mathcal{S}_{C_i}^x]$\;
}
$R\leftarrow Sort(C)$\Comment{\textrm{\small in descending order of rating}[$C_i$]}\;
}

\end{algorithm}
\subsubsection{Results and Analysis}
Our intuition behind R-Rec was to give more preference to user assigned ratings followed by user assigned tags. This method generated a ranked list of candidate suggestion which was then evaluated using the TREC evaluation metrics. The results are presented in Table \ref{tab:rrec}.
\begin{table}[!h]
\centering
\begin{tabular}{|c|c|c|}\hline
&P@5 & MRR\\ \hline \hline
Median of TREC-CS 2015 \cite{TREC15}& 0.5090 & 0.6716\\ \hline
Best system at TREC-CS 2015 \cite{Ali15} & 0.5858 & 0.7404\\ \hline
\textbf{R-Rec}& \textbf{0.5886}& \textbf{0.7461}\\ \hline
\end{tabular}
\caption{Comparison of R-Rec with TREC 2015 results}
\label{tab:rrec}
\end{table}
As can be seen, our approach outperforms the best system that took part in TREC 2015 Contextual Suggestion track. The highlight of our technique is its simplicity and elegance in capturing the intuition behind user's rating and tagging. The main reason why we lagged behind in our earlier approaches was due to the fact that we over enriched the tags for user preferences which might have led to a diluted overall score of a suggestion in the ranked list. Also noteworthy, is the fact that our hypothesis that ratings capture more essence of user's preference than tags is corroborated by the significant improvement in results.

\section{Conclusions}
Recommender systems has pervaded across all internet based e-commerce applications. Travel and tourism industry is no exception. In light of this, contextual suggestion aims at providing recommendations to users with past personal preferences, attractions or points-of-interest that the user might be interested in visiting. TREC Contextual Suggestion track investigates and explores this particular user-preference oriented problem. Previous approaches have tried various machine learning based approaches to address this problem while differing in methodology. In this paper, we show by means of thorough experimentation how a rule-based technique \textit{R-Rec} which takes into account both user ratings and tags can outperform all other approaches. Since each and every user has particular needs, we prove that in such cases individual preferences must be given more priority than collaborative choices. Our system was evaluated using two established metrics Precision at 5 (P@5) and MRR where it performed better than others. 
\bibliographystyle{abbrv}
\bibliography{context_sugg}  % sigproc.bib is the name of the Bibliography in this case
% ACM needs 'a single self-contained file'!
%\balancecolumns
%\balancecolumns % GM June 2007
% That's all folks!

\balancecolumns
% That's all folks!
\end{document}